\documentclass[10pt,a4paper,twoside]{article}

\pdfoutput=1

\usepackage{graphicx}
\usepackage{rotating}
\usepackage{tikz}

\usepackage{graphicx}
\usepackage{pstricks}
\usepackage{dcolumn}
\usepackage{bm}
\usepackage{algorithmic,pstricks}
\usepackage{algorithm}
\usepackage{amsmath}
\usepackage{amssymb,latexsym,amsfonts}
\usepackage{tikz,slashbox}
\usepackage{subfigure}

\title{Rhythmic behavior in a two-population mean field Ising model}

\author{Francesca Collet\thanks{Institute of Applied Mathematics, Delft University of Technology, Mekelweg 4, 2628 CD Delft (The Netherlands); e-mail addresses: F.Collet-1@tudelft.nl} \and Marco Formentin\thanks{Dipartimento di Matematica, Universit\`a degli Studi di Padova, Via Trieste 63, 35121 Padova (Italy); e-mail address: marco.formentin@rub.de} \and Daniele Tovazzi\thanks{Dipartimento di Matematica, Universit\`a degli Studi di Padova, Via Trieste 63, 35121 Padova (Italy); e-mail address: tovazzi@math.unipd.it}} 

\begin{document}

\maketitle

\begin{abstract}
\noindent Many real systems comprised of a large number of interacting components, as for instance neural networks , may exhibit collective periodic behavior even though single components have no natural tendency to behave periodically. Macroscopic oscillations are indeed one of the most common self-organized behavior observed in living systems. In the present paper we  study some dynamical features of a two-population generalization of the mean field Ising model with the scope of investigating simple mechanisms capable to generate rhythm in large groups of interacting individuals. We show that the system may undergo a transition from a disordered phase, where the magnetization of each population fluctuates closely around zero, to a phase in which they both display a macroscopic regular rhythm. In particular, there exists a region in the parameter space where having two groups of spins with inter- and intra-population interactions of different strengths suffices for the emergence of a robust periodic behavior.

\vspace{0.3cm}

\end{abstract}

	

\section{Introduction}

\noindent Living systems are characterized by the emergence of recurrent dynamical patterns at all scales of magnitude. Self-organized collective behaviors are observed both in large communities of microscopic components - like neural oscillations and gene network activity - as well as on larger levels - as predator-prey equilibria, applauding audiences and flock of birds to name fews. In particular, collective periodic behaviors are among the most commonly observed ways of self-organization in biology, ecology and socioeconomics \cite{ermentrout2010mathematical,turchin1992complex,chen2008nonlinear,weidlich2012concepts}. The attempt of modeling such complex systems leads naturally to consider large families of microscopic identical units. Complexity and self-organization then arise on a macroscopic scale from the dynamics of these minimal components that evolve coupled by interaction terms.\\
Within this scenario, we are interested in particle systems whose macroscopic observables oscillate between different ordered phases. In these models, microscopic units neither have tendency to behave periodically on their own -- in contrast to Kuramoto rotators \cite{Kur75} -- nor are subject to a periodic forcing. Nevertheless  the particles organize to produce a very regular motion perceived only macroscopically: a collective self-sustained rhythm. Various stylized models have been proposed to capture the essence of this phenomenon, 
but in most of them rigorous results are hard to obtain, as the study ends up in looking for stable attractors of nonlinear infinite dimensional dynamical systems \cite{LG-ONS-G04,GiPo15}. Analytically tractable models can be obtained by considering mean field interactions. Recently, existence of periodic collective behaviors has been proven for some classes of mean field systems derived as perturbation of classical reversible ferromagnetic models by adding a dissipation term \cite{dfr,DaPGiRe14,CoDaPFo15}. 
Dissipation dumps the influence of interaction when no transition occurs for a long time. The simplest spin system within this class is the dissipative mean field Ising model proposed in \cite{dfr,DaPGiRe14}.  Coupled diffusions with dissipation have been considered in \cite{CoDaPFo15}. Besides dissipation, delay in the interactions may also produce rhythmic behavior in mean field systems as highlighted in \cite{DiLo} for interacting Hawkes processes and in \cite{Tou} for  spin glass models. Indeed theoretical models based on mean field interacting spin systems, although simplistic, are able to show a good qualitative description of cooperative macroscopic behavior in self-organizing systems. In the last decades, for this reason and their analytical tractability, they have also been applied in social sciences \cite{BrDu01, CoDaPSa10, CoGh07}, finance \cite{FrBa08, DPRST09}, chemistry \cite{DBAgBaBu12} and ecology \cite{VoBaHuMa03, BDPFFM14}.\\ 
An interesting family, which has naturally emerged in applications, is a multi-species extension of the mean field Ising model. The possibility of taking account for several kinds of magnetic spins is a peculiar feature that may be relevant to capture diverse phenomena from magnetism in anisotropic materials  to social issues. A two-population version of the Curie-Weiss model was introduced in the 50s to mimic the phase transition undergone by metamagnets \cite{ScDa97}. Recently, it has been receiving a renewed attention due to its ability to describe the large scale behavior of socio-economic systems, such as cultural coexistence, immigration and integration \cite{CoGh07,CoGaMe08, GaBaCo09,BCSV14}. Multi-populated non-interacting spin models are the cornerstone of McFadden discrete choice theory \cite{McF01}. The extension of the discrete choice theory to the interacting, and more realistic, case has been done in \cite{CoGh07} and represents an important step toward the understanding of collective behaviors in societies. From an equilibrium viewpoint, the investigation of the two-species model introduced in \cite{CoGh07} has been pursued at a mathematical level in \cite{GaCo08}, where the thermodynamic limit has been rigorously obtained. In the present paper, following the work started in \cite{Col14}, we continue the analysis of the dynamical features of this two-population generalization of the mean field Ising spin system.\\
Inspired by previous works with applications in biology \cite{FeFoNe09}, neurosciences \cite{DiLo} and socioeconomics \cite{Tou}, our purpose is to investigate mechanisms that enhance rhythmic behavior. 
The goal of the paper is not to have a comprehensive study of the dynamics but rather to show the onset of regular behavior. Our main finding indicates that having two groups of spins with possibly different size and different inter- and intra-population interactions suffices for the emergence of macroscopic oscillations. Additional mechanisms as dissipated or delayed interactions are not necessary. However delay may produce periodic behavior in interaction network configurations where otherwise absent. In our approach transition to rhythm is detected in the thermodynamic limit via the presence of a Hopf bifurcation. Stable limit cycles may also emerge from non local bifurcations \cite{CoDaPFo15}, but there is no numerical evidence it could be the case for the class of mean field systems considered here.\\
\section{Model}

\begin{figure}[hb!] \centering
\begin{tikzpicture}
\draw (1.,1.5) circle (1.3cm);
\draw (5.,1.5) circle (1.3cm);
\node (3) at (1.,1.5) {$\mbox{Population $I_1$}$};
\node (4) at (5.,1.5) {$\mbox{Population $I_2$}$};
\draw [-latex,orange,thick, bend right] (2.2,1.) to (3.8,1.);
\draw [-latex,cyan,thick, bend right] (3.8,2) to (2.2,2);
\draw [latex-latex,blue,thick, bend right]  (5.5,.9) to (4.65,0.55);
\draw [latex-latex,red,thick, bend left]  (.8,2.5) to (1.75,2.1);
\draw (5.15,0.6) node {$J_{22}$};
\draw (1.3,2.2) node {$J_{11}$};
\draw (2.9,0.58) node {$J_{12}$};
\draw (2.9,2.44) node {$J_{21}$};
\shade[ball color=blue]  (4.1,1.) circle (.1cm);
\shade[ball color=blue]  (4.6,0.5) circle (.1cm);
\shade[ball color=blue]  (4.1,1.) circle (.1cm);
\shade[ball color=blue]  (5.6,0.9) circle (.1cm);
\shade[ball color=blue]  (5.,1.1) circle (.1cm);
\shade[ball color=blue]  (4.1,1.) circle (.1cm);
\shade[ball color=blue]  (4.7,2.5) circle (.1cm);
\shade[ball color=blue]  (4.9,1.9) circle (.1cm);
\shade[ball color=blue]  (5.1,2.3) circle (.1cm);
\shade[ball color=blue]  (5.9,2.1) circle (.1cm);
\shade[ball color=blue]  (4.1,2.) circle (.1cm);
\shade[ball color=red]  (.4,0.7) circle (.1cm);
\shade[ball color=red]  (.8,0.5) circle (.1cm);
\shade[ball color=red]  (.1,1.) circle (.1cm);
\shade[ball color=red]  (1.9,0.9) circle (.1cm);
\shade[ball color=red]  (1.,1.1) circle (.1cm);
\shade[ball color=red]  (1.1,.8) circle (.1cm);
\shade[ball color=red]  (.7,2.5) circle (.1cm);
\shade[ball color=red]  (1.,1.8) circle (.1cm);
\shade[ball color=red]  (.3,2.3) circle (.1cm);
\shade[ball color=red]  (.7,2.1) circle (.1cm);
\shade[ball color=red]  (1.8,2.) circle (.1cm);
\end{tikzpicture}
\caption{\footnotesize{A schematic representation of the interaction network for a bipartite Curie-Weiss model.  Spins are divided into two populations $I_1$ and $I_2$. Within $I_1$ (resp. $I_2$) particles feel a mean field interaction with coupling $J_{11}$ (resp. $J_{22}$). Beside, population $I_1$ (resp. $I_2$) influences the dynamics of the other group through its magnetization with strength $J_{12}$ (resp. $J_{21}$).}}
\label{CW}
\end{figure}
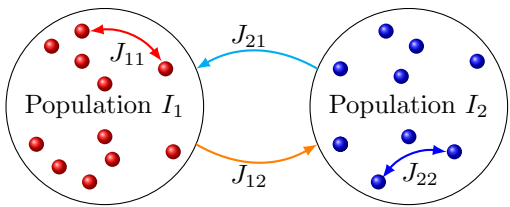

\noindent The two-population Curie-Weiss model is a spin system where on the complete graph two types of spins are present. Particles are differentiated by their mutual interactions: there are two \emph{intra-group} interactions, tuning how strongly sites in the same group feel each other, and two \emph{inter-group} interactions, giving the magnitude of the influence between particles of distinct populations.
Let $S=\{-1,+1\}$ be the state space of a single spin variable and  let $\sigma=(\sigma_j)_{j=1}^N \in S^N$  be the $N$-site configuration. We divide the whole system of size $N$ into two disjoint subsystems of sizes $N_1$ and $N_2$ respectively. Let $I_1$ (resp. $I_2$) be the set of sites belonging to the first (resp. second) subsystem. We have $\text{card} (I_1)  = N_1$ and $\text{card}(I_2) = N_2$, with $N_1 + N_2 = N$. To fix notation, let $1, 2, \dots, N_1$ be the indices corresponding to particles in population $I_1$ and $N_1+1, N_1+2, \dots, N$ those of particles in population $I_2$, so that
\[ 
\begin{array}{rc|c}
& \mbox{\scriptsize Population $I_1$} & \mbox{\scriptsize Population $I_2$} \\
\sigma = & ( \sigma_1, \sigma_2, \dots, \sigma_{N_1} & \sigma_{N_1+1}, \sigma_{N_1+2}, \dots, \sigma_{N} ). 
\end{array}
\] 
Given two spins, their mutual interaction
depends on the subsystems they belong to.
In our setting $J_{11}$ and   $J_{22}$ tune the interaction within sites of the same subsystem; whereas, $J_{12}$ and $J_{21}$ 
control the coupling strength between spins located in different subsystems (see Fig. \ref{CW} for a schematic representation). All the interactions can be either positive or negative allowing both ferromagnetic and antiferromagnetic interactions.\\

We want to define two different Markovian dynamics in this setting. We need few notation. Let us denote by%
\[
m_{N_i} (t) := \frac{1}{N_i} \sum_{j \in I_i} \sigma_j(t) 
\]
the magnetization of population $I_i$ ($i=1,2$) at time~$t$. Moreover, if  $\alpha := N_1/N$ is the proportion of sites belonging to the first group, we introduce the functions
\begin{align}
R_1 \left(x_1,x_2  \right) &= \alpha J_{11} x_1+ (1-\alpha) J_{12} x_2 \label{Def:R1} \\[0.2cm]
R_2 \left(x_1,x_2  \right) &= (1-\alpha) J_{22} x_2  + \alpha J_{21} x_1\,. \label{Def:R2}
\end{align}
The $R_i$'s are comprised of two terms: the first one tells how strong sites in the same population interact, while the second encodes the way one population influences the other. \\
We are now ready to describe the two dynamics we are interested in. For reasons that will be clear in a moment, throughout the paper we will refer to these dynamics as ``without delay'' and ``with delay''.\\

\noindent {\bf Microscopic dynamics without delay.}  Let $\sigma^i$ denote the configuration obtained from $\sigma$ by flipping the $i$-th spin. At any time $t$ the system may experience a transition whose rate depends on the magnetization vector at \emph{time $t$ only}.\\ 
The transition $\sigma \longrightarrow \sigma^i$ occurs  at rate 
\begin{equation}\label{without:delay:dyn}\tag{wD}
\left\{
\begin{array}{lcl}
 e^{-\sigma_i\:R_1 \left(m_{N_1} (t),m_{N_2} (t)   \right)}, && \mbox{ if } i \in I_1\\
e^{-\sigma_i\: R_2 \left( m_{N_1} (t),m_{N_2} (t)  \right)}, && \mbox{ if } i \in I_2 \,.
\end{array}
\right.
\end{equation}
These are standard Glauber Markovian dynamics where, for any small $\delta > 0$, the transition probability $P \{ \sigma (t+\delta) \vert \sigma(s), s \leq t\}$ depends only on the configuration at time $t$, i.e. $P \{ \sigma (t+\delta) \vert \sigma(s), s \leq t\} = P \{ \sigma (t+\delta) \vert \sigma(t) \}$.\\

\noindent {\bf Microscopic dynamics with delay.} For us it will be of interest also a second type of dynamics in which a delay kernel acts on the inter-group interactions. At any time $t$ the influence of each population on the other is given by an average over the magnetization trajectory \emph{up to time $t$}, weighted through a delay kernel.\\ 
The transition $\sigma \longrightarrow \sigma^i$ occurs  at rate 
\begin{equation}\label{with:delay:dyn}\tag{D}
\left\{
\begin{array}{lcl}
 e^{-\sigma_i\:R_1 \left(m_{N_1} (t),\,\gamma^{(n)}_{N_2}(t)   \right)}, && \mbox{ if } i \in I_1\\
e^{-\sigma_i\: R_2 \left( \gamma^{(n)}_{N_1}(t),\,m_{N_2} (t)  \right)}, && \mbox{ if } i \in I_2 \,,
\end{array}
\right.
\end{equation}
where, for $n \in \mathbb{N}$ and $k \in \mathbb{N} \setminus \{0\}$, we define
\begin{equation*}
\gamma^{(n)}_{N_i}(t) = \int_0^t \! \frac{(t-s)^n}{n!} k^{n+1} e^{-k(t-s)}\:m_{N_i} (s)\, ds, 
\end{equation*}
for $i=1,2$. {The delay kernel is in form of Erlang distribution. The parameter $n$ is related to the shape of the bump of the function; whereas, $k$ tunes how sharp and close to time $t$ the peak is. In particular, for any fixed $n \in \mathbb{N}$ and large $k$, the kernel has a sharp peak around $s\simeq t$, where the maximum is attained. As a consequence, in the limit as $k$ goes to infinity, only values of the magnetizations close to $m_{N_i}(t)$, for $i = 1, 2$, enter the dynamics and the rates \eqref{with:delay:dyn} approach the rates \eqref{without:delay:dyn}. This indicates that the addition of delay is relevant for small values of the parameter $k$.}


\section{Results}

\noindent We want to characterize the infinite volume limits of the dynamics \eqref{without:delay:dyn} and \eqref{with:delay:dyn} described above. The strategy is to determine a suitable Markov process, whose dynamics can be derived from the original microscopic dynamics, and then apply standard techniques of convergence of generators to get weak convergence to the thermodynamic limiting evolution \cite[Corollary 8.7, Chapter 4]{EtKu86}. This machinery has been applied in detail in \cite{BDPFFM14} for the treatment of a similar system.\\
Note that the limit as $N$ goes to infinity must be taken in such a way the proportions $\alpha$ and $1-\alpha$ of the two groups remain constant. In the sequel  we will write $m_i(\cdot)$ for the infinite volume limit of $m_{N_i}(\cdot)$. Analogously $\gamma^{(n)}_{i}(\cdot)$ will be the limit of $\gamma^{(n)}_{N_i}(\cdot)$. \\

\noindent {\bf Macroscopic dynamics without delay.} The dynamics \eqref{without:delay:dyn} for configurations induce a Markovian evolution on the magnetization vector $(m_{N_1}(t),m_{N_2}(t))$. As $N \longrightarrow \infty$, the process $(m_{N_1}(t),m_{N_2}(t))_{t \geq 0}$ weakly converges to the solution of the system of ordinary differential equations
\begin{equation}\label{MKV}\tag{MwD}
\begin{array}{lcl}
\dot{m}_1(t) & = & 2 \sinh \left[ R_1 \left( m_1(t), m_2(t) \right) \right] \\
&& -  2 m_1(t) \cosh \left[ R_1 \left( m_1(t), m_2(t) \right) \right] \\     
\\
\dot{m}_2(t) & = & 2 \sinh \left[ R_2 \left( m_1(t), m_2(t) \right) \right]\\
&& -  2 m_2(t) \cosh \left[ R_2 \left(m_1(t), m_2(t) \right) \right] \,.
\end{array}
\end{equation}

\noindent {\bf Macroscopic dynamics with delay.} In this case the magnetization vector in itself does not inherit Markovianity from \eqref{with:delay:dyn}. To get a Markovian evolution for macroscopic observables we have to consider the process 
\[
\left( m_{N_1}(t), m_{N_2}(t), \left( \gamma_{N_1}^{(j)}(t) \right)_{j=0}^n, \left( \gamma_{N_2}^{(j)}(t) \right)_{j=0}^n\right)_{t \geq 0},
\]  
that, as $N \longrightarrow \infty$, weakly converges to the solution of the following system of ordinary differential equations
\begin{equation}\label{MKVwithD}\tag{MD}
\begin{array}{lcl}
\dot{m}_1(t) & = & 2 \sinh \left[ R_1 \left(m_{1} (t),\,\gamma^{(n)}_2(t) \right) \right] \\
& & -  2 m_1(t) \cosh \left[ R_1 \left(m_{1} (t),\,\gamma^{(n)}_2(t)   \right) \right] \\     
\\
\dot{m}_2(t) & = & 2 \sinh \left[ R_2 \left( \gamma^{(n)}_1(t),\, m_2(t) \right) \right]\\
& & -  2 m_2(t) \cosh \left[ R_2 \left( \gamma^{(n)}_1(t),\, m_2(t) \right) \right]\\
\\
\dot{\gamma}^{(0)}_1(t) & = &k \left[-\gamma^{(0)}_1(t) + m_1(t) \right]\\
\\
\dot{\gamma}^{(n)}_1(t) & = &k \left[ -\gamma^{(n)}_1(t)+\gamma^{(n-1)}_1(t) \right] \mbox{, for $n>0$}\\
\\
\dot{\gamma}^{(0)}_2(t) & = &k \left[ -\gamma^{(0)}_2(t) + m_2(t) \right]\\
\\
\dot{\gamma}^{(n)}_2(t) & = &k \left[ -\gamma^{(n)}_2(t)+\gamma^{(n-1)}_2(t) \right] \mbox{, for $n>0$}\,.
\end{array}
\end{equation}
We remark that introducing delay through a kernel (idea borrowed from \cite{DiLo}) leads to a finite dimensional macroscopic dynamics. In contrast, if instead of $\gamma_{N_i}^{(n)}(t)$ we choose $\overline{\gamma}_{N_i} = m_{N_i} (t - \tau)$, with fixed $\tau > 0$ (\emph{delayed rates}), the limiting dynamics are infinite dimensional. A detailed analysis of  a mean field spin system with delayed rates is given in \cite{Tou}. In addition therein, the author considers a spatial model where both the interaction and the delay depend on respective locations of sites.\\

{
\noindent 
It is evident from equation \eqref{MKVwithD} that, for \emph{large} $k$, there is a separation of time scales between the evolutions of the $m$-variables and the $\gamma$-variables: the latter relax to their equilibrium point much faster than the former. By applying the center manifold reduction \cite[Theorem 5.2]{Kuz04}, it is possible to reduce the $(2n+4)$-dimensional dynamical system \eqref{MKVwithD} to a planar dynamical system, describing the $O(1)$ evolution of $(m_1,m_2)$ on the center manifold 
\[
\gamma_1^{(0)} = \cdots = \gamma_1^{(n)} = m_1 \mbox{ and } \gamma_2^{(0)} = \cdots = \gamma_2^{(n)} = m_2.
\]
Thus we can neglect the equations for the $\gamma$-variables and consider only the dynamics of the $m$-variables after having substituted the stationary values for the $\gamma$'s. It follows that, for large $k$, the reduction of \eqref{MKVwithD} coincides with the macroscopic dynamics without delay \eqref{MKV}. This calls for a few observations:
\begin{itemize}
\item the proximity between the macroscopic evolutions \eqref{MKV} and \eqref{MKVwithD} we obtain for large $k$ is not so unexpected. It somehow reflects the fact that the microscopic dynamics \eqref{without:delay:dyn} and \eqref{with:delay:dyn} are close to each other as $k$ goes to infinity. Indeed, since $\gamma_{N_i}^{(n)}(t) \xrightarrow[]{k \to + \infty} m_{N_i}(t)$, for all $n \in \mathbb{N}$ and $t \in \mathbb{R}^+$, there is convergence of the transition rates \eqref{with:delay:dyn} towards \eqref{without:delay:dyn} and, as a consequence, the two stochastic processes become close the one to the other.
\item the proximity between the dynamics \eqref{MKV} and \eqref{MKVwithD} holds true \emph{only in the large $k$ limit}. It will be clear in the next section that, for small $k$, the qualitative behaviors of the two systems may significantly differ. Delay may produce periodic behavior in interaction network configurations where otherwise absent.
\end{itemize}
}

\begin{figure}[ht!]  
\centering%
\includegraphics[angle=90,width=1\textwidth]{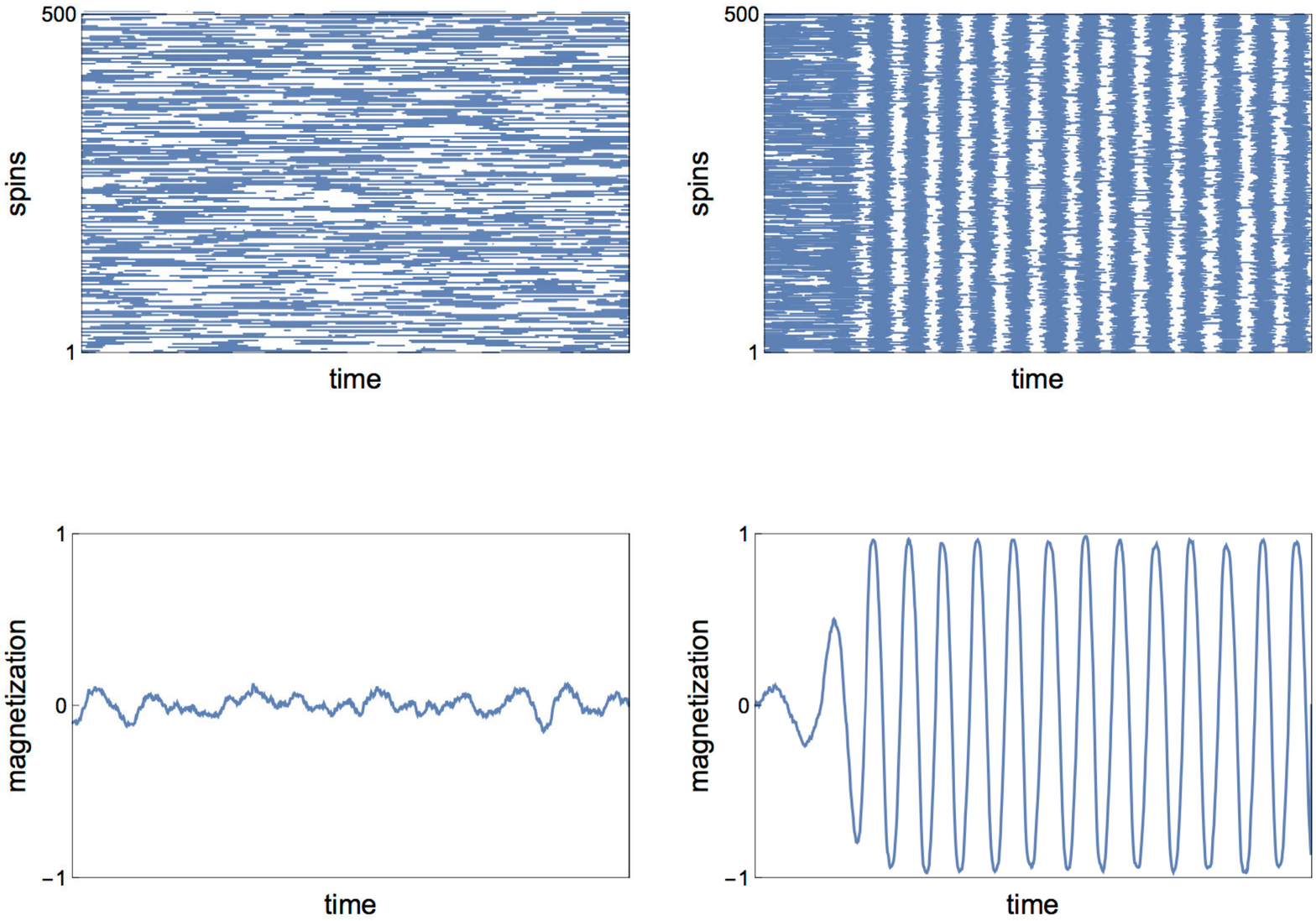}
\caption{\footnotesize{Transition from disordered behavior (on the left) to collective rhythm (on the right) for the spin system \eqref{without:delay:dyn}. Simulations have been run with $N=1000$, $\alpha=1/2$, $J_{12}=-6$, $J_{21}=5$ and $J_{11}=J_{22}=0.5$ on the left and $J_{11}=J_{22}=3$ on the right. The top row shows the time evolution of all the spins belonging to population $I_1$. Spins are labelled from 1 to 500 on the $y$-axis. Blue spots represent $+1$ spins; whereas, white spots stand for $-1$. In the bottom line the corresponding evolution for the magnetization is depicted.}
}
\label{fig:like_neurons}
\end{figure}

\noindent {\bf Transition from disorder to rhythm.} We want to detect the transition from a disordered behavior, where $m_{N_1}(\cdot)$ and $m_{N_2}(\cdot)$ fluctuate around zero, to a collective rhythmic behavior in which we have periodic motion of the magnetizations (see Fig. \ref{fig:like_neurons}). To this aim we consider the limiting evolutions \eqref{MKV} and \eqref{MKVwithD} and  we look for the presence of a Hopf bifurcation. 
Recall that a (supercritial) Hopf bifurcation occurs when a stable periodic orbit arises from an equilibrium point as, at some critical values of the parameters, it loses stability. Such a bifurcation can be detected when a pair of complex eigenvalues of the linearized system around the equilibrium crosses the imaginary axis \cite[Theorem 
2, Chapter 4.4]{Per01}.\\

\begin{figure}[ht!]  
\centering%
\includegraphics[angle=90,width=1\textwidth]{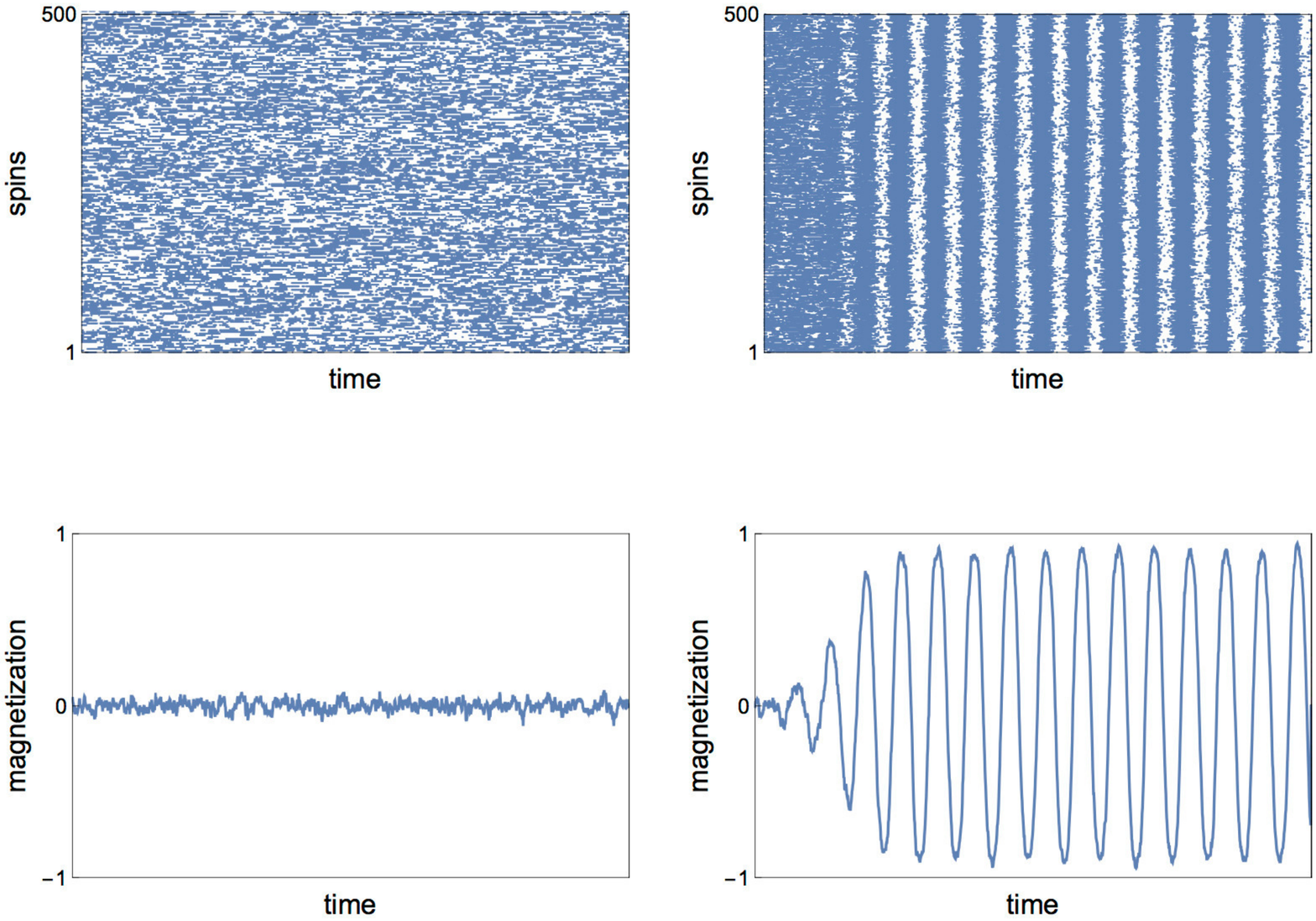}
\caption{\footnotesize{Transition from disordered behavior (on the left) to collective rhythm (on the right) for the spin system \eqref{with:delay:dyn} when both the intra-group interactions are negative (i.e. $J_{11}, J_{22}  < 0$). Simulations have been run with $N=1000$, $\alpha=1/2$, $n=2$, $J_{12}=5$, $J_{21}=-6$ and $J_{11}=J_{22}=-4$ with $k=6$ on the left and $J_{11}=J_{22}=-1$ with $k=3$ on the right. Recall that the value of $k$ depends on $J_{11}$ and $J_{22}$ through our simplifying assumption \eqref{relation:k:J}. The top row shows the time evolution of all the spins belonging to population $I_1$. Spins are labelled from 1 to 500 on the $y$-axis. Blue spots represent $+1$ spins; whereas, white spots stand for $-1$. In the bottom line the corresponding evolution for the magnetization is depicted.}}
\label{fig:like_neurons_bis}
\end{figure}

\noindent We start by considering the set of ordinary differential equations given by  \eqref{MKV}.  It is immediate to verify that the origin is an equilibrium for all values of the parameters. Therefore we analyze the spectrum of the linearization of the dynamics \eqref{MKV} around $(0,0)$ to understand if there exist parameter values for which the origin loses stability whenever a pair of pure imaginary conjugate eigenvalues appear.  The characteristic polynomial of the linearized system reads
\begin{align*}
P(\lambda) &= \lambda^2-\lambda(2\alpha J_{11}+2(1-\alpha)J_{22}-4)\\
&+4(\alpha J_{11}-1)\left((1-\alpha)J_{22}-1\right)\\
&-4\alpha(1-\alpha)J_{12}J_{21}.
\end{align*}
It follows that a Hopf bifurcation occurs if and only if both the conditions 
\begin{equation}\label{HopfBifurcation}
\begin{array}{lcl}
\alpha J_{11}=2-(1-\alpha) J_{22}\\[0.2cm]
\left((1-\alpha)J_{22}-1\right)^2+\alpha(1-\alpha)J_{12}J_{21}<0
\end{array}
\end{equation}
are satisfied. In particular we have:
 \begin{itemize}
 \item If $J_{11}$, $J_{22} \leq 0$, the equality in \eqref{HopfBifurcation} is never satisfied and thus system \eqref{MKV} never undergoes a Hopf bifurcation.
 \item If $J_{12}J_{21} \geq 0$, the inequality in \eqref{HopfBifurcation} has no solution and then again it is impossible to find a Hopf bifurcation.
 \item In the set $\{J_{11},\, J_{22} \leq 0\}^c \cap \{J_{12}J_{21} \geq 0\}^c$ we can choose properly the values of the parameters to get a Hopf bifurcation.
 \end{itemize} 
Our aim is now to understand if, in those regions of the parameter space where \eqref{MKV} does not undergo a Hopf bifurcation, we may produce a transition to periodic motion by adding delay in the dynamics. In this respect we move to the analysis of \eqref{MKVwithD}. It is easy to see that whenever a Hopf bifurcation is present for \eqref{MKV}, the same holds also for \eqref{MKVwithD}. Delay may of course change the critical value at which the bifurcation occurs, but not its presence. Moreover, by adding delay we can induce rhythmicity in a subspace of the phase $J_{11}$, $J_{22} \leq 0$, where periodic orbits were absent for \eqref{MKV}.\\ 
We consider the set of mean field equations in \eqref{MKVwithD} with $J_{11}$, $J_{22} < 0$ and $J_{12}J_{21}<0$. We then linearize the dynamics around the null $2n+4$-dimensional vector, which indeed is fixed point of \eqref{MKVwithD} for all parameter values,  and we study the spectrum. We remark that, since dealing with a non-planar dynamical system, to detect a supercritical Hopf bifurcation it does not suffice looking for a pair of pure imaginary conjugate eigenvalues; in addition, it is necessary to check that all the $2n+2$ remainings have negative real part.\\
To simplify computations assume 
$$ 2(\alpha J_{11}-1)=-k \quad \mbox{ and } \quad 2\left((1-\alpha)J_{22}-1\right)=-k,$$
so that 
\begin{equation}\label{relation:k:J}
\alpha J_{11}=(1-\alpha) J_{22}=-\frac{k}{2}+1,
\end{equation}
with $k > 2$. The constant $k$ we are using here is the same appearing in the definition of the delay kernel. Let us denote by $x_j$, with $j=0, 1, \ldots, 2n+3$, the $j-$th eigenvalue of the linearization of \eqref{MKVwithD} around the null solution. For $j=0, 1, \ldots, 2n+3$, we get
\[
x_j = - k - \left|A\right|^{\frac{1}{2n+4}} \, k^{\frac{n+1}{n+2}} \, \exp\left\{ i\frac{(2j+1)\pi}{2n+4} \right\},
\] 
where $A=4\:\alpha(1-\alpha) J_{12}J_{21}<0$. Therefore, for
\begin{equation}\label{HopfBifurcationD} k= \left|A\right|^{\frac{1}{2}} \, \left[ \cos \left(\frac{\pi}{2n+4} \right) \right]^{n+2} \end{equation}
a Hopf bifurcation occurs, as $x_{n+1}$ and $x_{n+2}$ are the two first eigenvalues passing the imaginary axis with positive derivative. See Fig. \ref{fig:like_neurons_bis}.\\

\noindent {We qualitatively summarize our findings in Table~\ref{tab:table1}. The table gives information about the possible emergence of macroscopic oscillations for the interaction network configurations  depicted in the left column.  Given a type of  interaction network, the corresponding dynamical systems \eqref{MKV} (central column) and \eqref{MKVwithD} (right column) may or may not exhibit rhythmic behavior. When writing  they do, we mean that there exists a choice of the parameters 
(satisfying \eqref{HopfBifurcation} for \eqref{MKV}  and \eqref{HopfBifurcationD} for \eqref{MKVwithD})  for  which a Hopf bifurcation occurs at the origin.
We observe that the parameter $n$ does not influence the presence/absence of periodic behavior, but simply modifies the threshold value for the phase transition.}
Referring to the table, notice that only when both the intra-group interactions are negative (i.e. $J_{11}, J_{22}  < 0$) delay is needed to enhance the transition to a periodic behavior for $(m_1,m_2)$. In all other cases a robust choice of the parameters is sufficient. In particular, delay is not necessary to create a limit cycle when $J_{11}, J_{12} < 0$ and $J_{22}, J_{21} > 0$ (third row in the table). 
The particle system constructed on the latter interaction network resembles one of the models introduced in \cite{Tou}, where however a fixed time-delay is present in the rates of transition. Our results indicate that it is not delay, but rather the asymmetry of the coupling strengths, the crucial feature to produce a collective rhythm. We believe that for the mean field spin-glass in \cite{Tou} the introduction of delay becomes necessary due to the fine choice of the interactions: $|J_{11}|=|J_{22}|=|J_{12}|=|J_{21}|=J$, with $J > 0$. 

\begin{table}[h!]
  \centering
  \caption{\footnotesize{Qualitative summary of the results. In the left column a schematic representation of the considered interaction network is displayed. The color convention for couplings is as in Fig.~\ref{CW}. For each interaction network we highlight the possibility of \emph{observing} or \emph{not observing} periodic behavior when considering the dynamics \eqref{MKV} (central column) or \eqref{MKVwithD} (right column). Notice that, in all cases except for one,  
delay is not necessary to produce rhythmic oscillations.}}
  \label{tab:table1}
  \begin{tabular}{c||c|c}
 &&\\
\backslashbox{Interactions}{Dynamics} &{{\emph{without} delay}}& {{\emph{with} delay}}\\
&&\\
  \hline
  \hline
\includegraphics[width=25mm,scale=0.3]{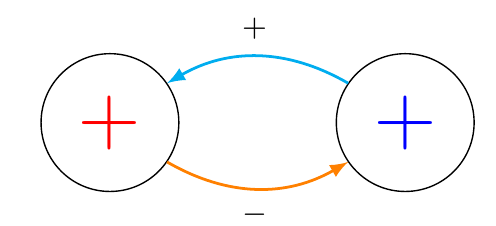}&\includegraphics[width=20mm,scale=0.3]{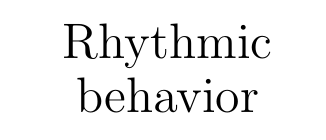}& \includegraphics[width=20mm,scale=0.3]{yes_bis}\\
    \hline
\includegraphics[width=25mm,scale=0.3]{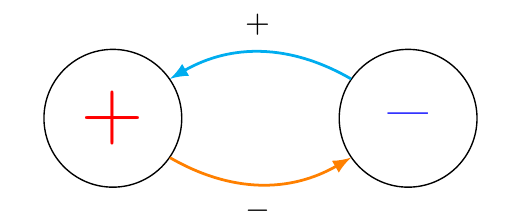}&\includegraphics[width=20mm,scale=0.3]{yes_bis}& \includegraphics[width=20mm,scale=0.3]{yes_bis}\\
    \hline
     \includegraphics[width=26.5mm,scale=0.3]{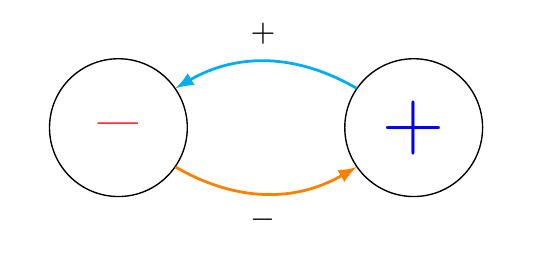}&\includegraphics[width=20mm,scale=0.3]{yes_bis}& \includegraphics[width=20mm,scale=0.3]{yes_bis}\\
    \hline
  \includegraphics[width=26.5mm,scale=0.3]{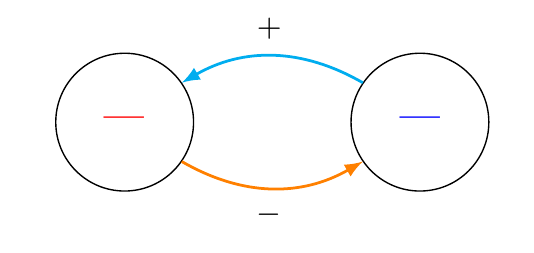}&\includegraphics[width=22mm,scale=0.35]{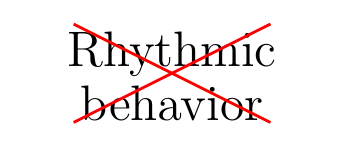}& \includegraphics[width=20mm,scale=0.3]{yes_bis}\\
\hline
  \end{tabular}
\end{table}
\section{Conclusions}

In the present paper we investigated the emergence of collective periodic behavior in a two-population generalization of the mean field Ising model. We analyzed the role of interaction network and delay in enhancing an oscillatory evolution for the magnetization vector. We were interested in showing that it is possible to induce a transition from a disordered phase, where $m_{N_1}$ and $m_{N_2}$ fluctuate closely around zero, to a phase in which they both display a macroscopic regular rhythm. In particular we have proven that a robust 
choice of the coupling constants and of the population sizes is sufficient for a limit cycle to arise. Moreover, in the case when the choice of the parameters does not suffice to favor the transition, delay may help in this respect (see Table~\ref{tab:table1}).\\
When considering the dynamics without delay, the mechanism behind the emergence of  
periodicity can be understood in the following terms. If the intra-population interaction strengths $J_{11}$ and $J_{22}$ are large enough, each single population can be seen as a macro-spin that under Glauber dynamics tends to its own rest state. However, as soon as the two populations are linked together within an interaction network with $J_{12}J_{21}<0$, they form a frustrated pair of macro-spins where the rest state of the first is not compatible with the rest position of the second. As a consequence the dynamics is not driven to a fixed equilibrium and continues oscillating.
This intuition suggests that the creation of a collective rhythm by splitting the particles in two groups differentiated by their mutual interactions is very much related
to the mean  field setting, conclusion that is supported
also by numerics. We ran simulations of a two-population version of the nearest-neighbor Ising model on a finite square lattice of side-length $N \gg 1$ (total number of spins of order $10^3$). 
\begin{itemize}
\item
Particles are randomly divided in two distinct groups: each site in the box is assigned to population $I_1$ (resp. $I_2$) with probability $\alpha$ (resp. $1-\alpha$), with $\alpha \in ]0,1[$.
\item
At any time $t$, the spin $\sigma_k$ flips at a rate of the form \eqref{without:delay:dyn} in which the mean-field magnetizations $m_{N_i}(t)$ $(i=1,2)$ are replaced by \emph{local magnetizations}
\[
\ell_i (h,t) = \sum_{\substack{j \sim h \\ j \in I_i}} \sigma_j(t) \qquad  (i=1, 2)
\]
where the sum is extended only to sites $j$ nearest-neighbors of $h$ (as the symbol $\sim$ is intended to mean).
\item
Periodic boundary conditions are considered. 
\end{itemize}
With diverse simulations we explored the parameter space and did not find any oscillatory evolution for the global magnetizations of the two groups. The reason is that in this setting the short-range of interaction destroys the macro-spin structure of each family of spins and does not allow for the creation of a frustrated macro-network. The addition of delay does not change the scenario. \\
To conclude recall that the results we obtained for the two-population Curie-Weiss model 
are derived in the limit as the number of particles goes to infinity and the passage from an incoherent to a coherent behavior of the magnetization vector is detected by the occurrence of a (supercritical) Hopf bifurcation. It is worth to mention that the presence of a stable limit cycle as attractor for the dynamics is a pure infinite volume effect. The finite $N$ system \eqref{without:delay:dyn} is an irreducible, time homogeneous Markov process and therefore it relaxes to a time-independent invariant distribution. As a consequence, periodicity turns out to be only a metastable state for the finite size system.\\

\textit{Acknowledgement.}  We thank Pierluigi~Contucci  and Paolo~Dai~Pra for useful discussions,  comments and suggestions. 
FC was supported by the Netherlands Organisation for Scientific Research (NWO) via \mbox{TOP-1} grant 613.001.552. MF acknowledges the University of Padova, Mathematics Department Senior Grant 795/2014 Prot.~59008 for financial support.

\bibliographystyle{abbrv}

\end{document}